 \documentclass[final,5p,times,twocolumn]{elsarticle}

\usepackage{amssymb}
\RequirePackage{amsmath}

\makeatletter
\def\Hline{%
\noalign{\ifnum0=`}\fi\hrule \@height 1.5pt \futurelet
\reserved@a\@xhline}
\makeatother
\journal{Frontier of Applied Plasma Technology}

\begin{document}

\begin{frontmatter}



\title{Two-dimensional Modeling of the Hall Thruster Discharge \\with Non-uniform Propellant Supply in Azimuth}


\author[label1]{Rei Kawashima\corref{cor1}}

\cortext[cor1]{Corresponding author.}
%\ead{kawashima[at]al.t.u-tokyo.ac.jp}
\address[label1]{Department of Aeronautics and Astronautics, The University of Tokyo, 7-3-1 Hongo, Bunkyo, Tokyo 113-8656, Japan}
\address[label2]{Department of Advanced Energy, The University of Tokyo, 5-1-5 Kashiwanoha, Kashiwa, Chiba 277-8561, Japan}

\author[label1]{Junhwi Bak}
%\ead{j.bak@al.t.u-tokyo.ac.jp}

\author[label1]{Kimiya Komurasaki}
%\ead{komurasaki@al.t.u-tokyo.ac.jp}

\author[label2]{Hiroyuki Koizumi}
%\ead{koizumi@al.t.u-tokyo.ac.jp}

\begin{abstract}
   A two-dimensional simulation is conducted to investigate the effect of cross-field electron transport enhancement in the Hall thruster discharge caused by the nonuniform propellant supply in azimuth. 
   The Hall thruster operation with azimuthally nonuniform propellant supply is expected to be a good test case to understand the influences of azimuthal plasma property distributions on the axial electron transport. 
   A particle-fluid hybrid model is developed with a two-dimensional magnetized electron fluid model incorporated with an empirical anomalous electron mobility model. 
   The calculation results indicate that the azimuthal electric field is generated in the cases of nonuniform propellant supply. 
   The azimuthal phase shift is observed between the ion number density and space potential at the downstream plume region. 
   Owing to this azimuthal out-of-phase and the azimuthal electric field, the cross-field electron transport in the axial direction can be enhanced.
\end{abstract}

\begin{keyword}
   plasma simulation \sep hybrid model \sep Hall thruster \sep E$\times$B device \sep electron transport

\end{keyword}
\end{frontmatter}

%%%%%%%%%%%%%%%%%%%%%%%%%%%%%%%%%%%%%%%%%%%%%%%%%%%%%%%%%%%%%

\section{Introduction}
	\label{sec:intro}
	Hall thruster is one of the efficient electric propulsion devices used for spacecrafts. 
	Hall thruster has an annular discharge channel with a crossed-field configuration where a radial magnetic field and an axial electric field are applied, and the Hall current is induced in the azimuthal (E$\times$B) direction. 
	Owing to the magnetic confinement of electrons, an efficient ionization with short discharge channel is enabled and an electric field is maintained even with the plasma density of $\sim10^{18}$ m$^{-3}$. 
	Research and development of Hall thruster systems have been competitively conducted in various countries including US, Europe, and Japan \cite{HamadaJSASS2017}, and recently highly efficient thrusters with thrust efficiencies exceeding 60\% have been developed. 
	The plasma flow simulation technologies for the Hall thrusters have also been advanced, and the approach to developing Hall thrusters is shifting to the phase of computer-aided-engineering. 
	A deep understanding on the physics of the magnetized plasma in Hall thrusters remains important for accurate numerical simulations and efficient thruster developments. 

	The cross-field electron transport is one of the issues that is not fully understood in the physics of Hall thruster discharge. 
	It is well known that the classical electron diffusion theory is not sufficient to account for the plasma property distributions and discharge currents obtained by experiments. 
	To cope with this issue, most of the numerical works of Hall thrusters adopt empirical electron transport models for the agreement of simulation results with experimental data. 
	The issue of the anomalous electron transport has been found in various cross-field devices, and the mechanism responsible for the enhanced electron transport has been investigated for decades. 
	One theoretical mechanism proposed for the anomalous electron transport is related to the low-frequency azimuthal plasma oscillations in 10 kHz--100 kHz, such as rotating spokes \cite{Keidar:2006aa}. 
	The azimuthal oscillation induces an azimuthal electric field $E_{\theta}$ that causes the $E_{\theta}\times B_{r}$ drift in electron motion in the axial direction. 
	The existence of $E_{\theta}$ alone does not enhance the net electron current in the axial direction if one considers the integral of the full azimuth. 
	However, if the electron density is also nonuniform and there is an out-of-phase between the electron density and space potential, the net electron current in the axial direction is changed. 
	In other words, the axial electron transport can be enhanced if the following two conditions are satisfied: 1) the presence of $E_{\rm \theta}$, and 2) the out-of-phase between electron density and space potential.
	
\begin{figure}[t]
		\begin{center}
			\includegraphics[width=70mm]{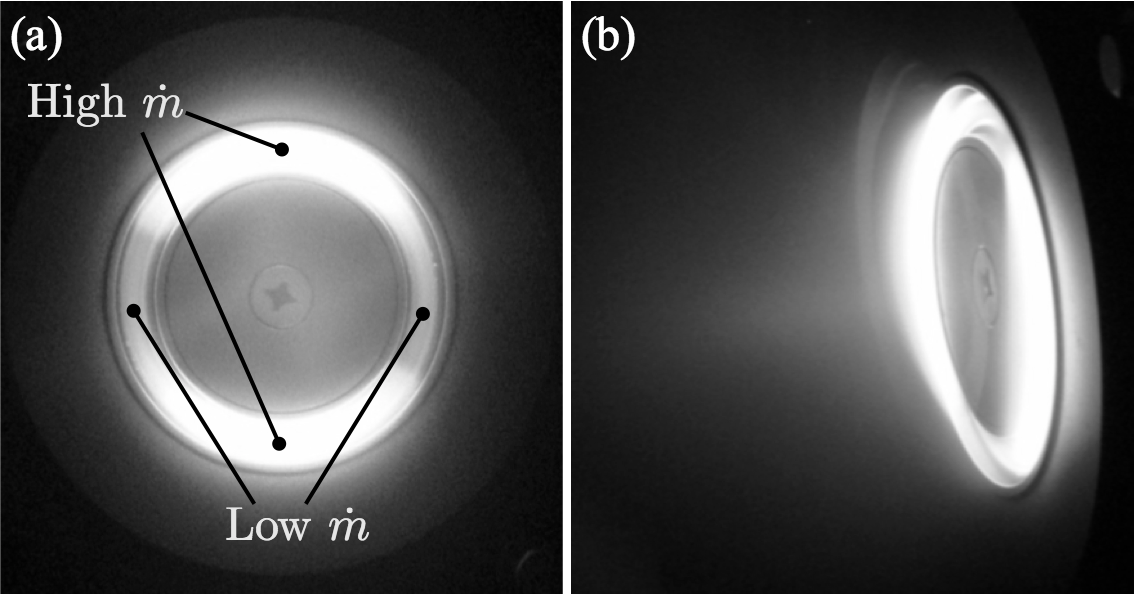}
		\end{center}
		\vspace{-4mm}
   	\caption{Thruster operation with azimuthal nonuniform propellant supply \cite{FukushimaIEPC2009}.
   	(a) front view, and (b) side view.}
   	\label{fig:intro}
   \end{figure}

	Another research that is related to the azimuthal physics and enhanced electron transport in Hall thruster is the addition of artificial nonuniformities in azimuth. 
	The azimuthally nonuniform propellant supply was tested with a 600 W-level thruster as shown in Fig. \ref{fig:intro} \cite{FukushimaIEPC2009}. 
	In this research, the hollow anode for the xenon gas supply was azimuthally divided into four sections, and the mass flow rates for the gas inlet port in these sections were differentiated. 
	The primary purpose of this nonuniform propellant supply was the mitigation of discharge current oscillation \cite{FukushimaIEPC2009}. 
	An additional important finding was that the discharge current was increased in the case of azimuthally nonuniform propellant supply whereas the ion beam current remained unchanged, compared with the case of uniform propellant supply \cite{BakIAPS2018}. 
	This means that the axial electron transport has been enhanced in the case of nonuniform propellant supply. 
	This research exemplifies the electron transport enhancement due to the azimuthally nonuniform plasma properties.
	
	The thruster operation with azimuthal nonuniform propellant supply has mainly two advantages for the investigation of cross-field electron transport.
	One is that the azimuthal plasma property distributions are steady so that the direct measurement of plasma property distributions is relatively easy.
	In the case of azimuthal plasma oscillations, the direct measurement is challenging since a high-speed measurement system is required to resolve the oscillations in 10 kHz--100 kHz.
	The other advantage is that the level of azimuthal nonuniformity is controllable by the differentiated mass flow rates. 
	In the operation of azimuthally nonuniform propellant supply, the electron current changes as a function of the parameter representing the level of azimuthal nonuniformity \cite{BakIAPS2018}. 
	This characteristic will enable a detailed parametric study on the relationship between azimuthal plasma property distributions and axial electron transport. 
	Therefore, the operation of a Hall thruster with azimuthally nonuniform propellant supply is expected to be a good test case to investigate the effects of azimuthal plasma properties on the cross-field electron transport.
	
	The objective of this research is the numerical simulation of a Hall thruster operation with azimuthally nonuniform propellant supply. 
	A numerical simulation based on a particle-fluid hybrid model is performed in the two dimensions corresponding to the E-field and E$\times$B directions. 
	We have checked the above-mentioned two conditions for the electron transport enhancement in the case of azimuthally nonuniform propellant supply: 1) the presence of $E_{\rm \theta}$, and 2) the out-of-phase between electron density and space potential. 
	In this paper, these points are qualitatively discussed.

%\clearpage	%%%%%%%%%%%%%%%%%%%%%%%%%%%%%%%%%%%%%%%%%%%%%%%%%%%%%%%%%%%%%%%%%%%%%%%%%%%%%%%

\section{Two-dimensional Hybrid Model}
	\subsection{Physical model and numerical methods}
	The physical model and computational methods for the two-dimensional hybrid model are essentially consistent with the ones proposed in Ref. \cite{Kawashima:2018ab}. 
	Ion and neutral particle flows are calculated by the two-dimensional-three-velocity (2D3V) particle-in-cell (PIC) method. 
	The collision frequencies are calculated by using the empirical reaction rate coefficients \cite{GoebelKatz2008}. 
	To simplify the calculation, only first-ionization of xenon is treated as ions, and interparticle collisions between heavy particles are neglected.

	The fundamental equations for electron fluid are the two-dimensional electron mass and momentum conservation equations in quasi-neutral plasmas, and one-dimensional (1D) energy conservation equation. 
	In the mass conservation equation, the electron number density is treated as a time-constant distribution by assuming the quasineutrality. 
	The momentum conservation is expressed by the drift-diffusion equation obtained by neglecting the electron inertia. 
	For a stable computation of magnetized electron fluids, the hyperbolic system approach using the pseudo-time advancement technique is employed. 
	The processes for deriving the hyperbolic system were presented elsewhere \cite{Kawashima201559}. 
	The space discretization is implemented by the second-order upwind method, which has been verified for the magnetized electron fluid calculation in the coordinate of E-field and E$\times$B directions \cite{Kawashima:2018aa}.
	
	A number of models have been proposed for accurate expression of the electron mobility in the orthogonal direction of magnetic lines of force. 
	The most standard model for the cross-field electron mobility $\mu_\perp$ may be to include both the classical and anomalous electron mobilities as follows:
	\begin{equation}
		\mu_\perp = \frac{\mu_{||}}{1+\left(\mu_{||}B\right)^2}+\frac{\alpha_{\rm B}}{16B},
	\end{equation}
	where $B$ is the magnetic flux density and $\mu_{||}$ is the electron mobility of nonmagnetized electrons. 
	The Bohm-type diffusion model is assumed for the anomalous electron mobility where the $\alpha_{\rm B}$ is an empirical Bohm diffusion coefficient. 
	Figure 2 shows the distribution of $\alpha_{\rm B}$ assumed in this study, as a function of $x$-position normalized by the channel length $L_{\rm c}$. 
	To select the value of $\alpha_{\rm B}$, the three-region model \cite{Hofer:2008aa} is referred in which $\alpha_{\rm B}$ is assumed to be 0.14, 0.02, and 1.0 at the anode, channel exit, and plume regions, respectively. 
	To obtain a smooth distribution of $\alpha_{\rm B}$, the Gaussian functions are used in the present model. 
	The equations for these functions can be found in Ref. \cite{Kawashima:2018ab}.

	\begin{figure}[b]
		\begin{center}
			\includegraphics[width=65mm]{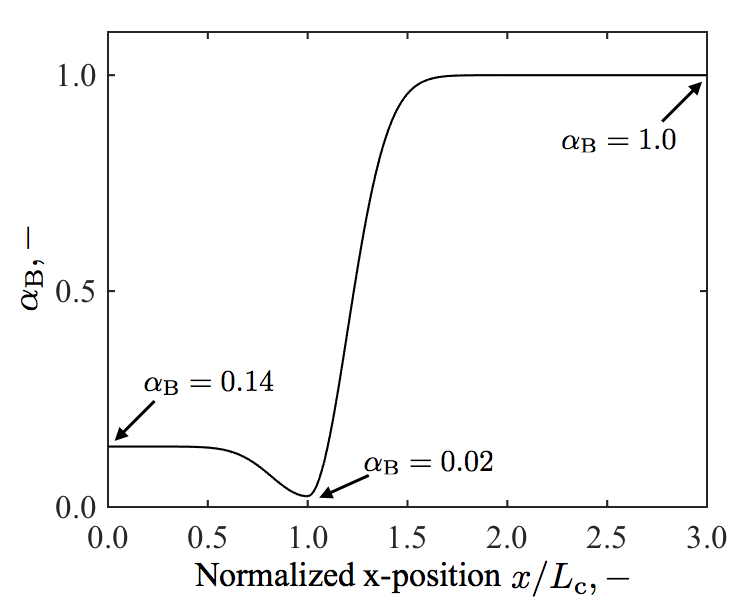}
		\end{center}
		\vspace{-4mm}
		\caption{Distribution of the Bohm diffusion coefficient in the E-field (axial) direction in the present model.}
		\label{fig:alp}
	\end{figure}

	\subsection{Calculation target: SPT-100}
	
	An annular Hall thruster similar to the SPT-100 thruster is assumed as the calculation target. The SPT-100 is a benchmark thruster in the Hall thruster community and several experimental and numerical researches have been conducted for this thruster. 
	The thruster configuration and operation condition of the simulated thruster are presented in Table \ref{tab:condition} \cite{Hofer:2007aa,ReidIEPC2015}.
	In the present 1D and 2D simulations that resolve E-field and E$\times$B directions, the radial effects including the radial gradients of plasma property and channel wall are neglected. 
	The magnetic field is assumed to be only in the radial direction, with a distribution in the E-field direction \cite{Mitrofanova}.
	
%\clearpage%%%%%%%%%%%%%%%%%%%%%%%%%%%%%%%%%%%%%%%%%%%%%%%%%%%%%%%%%%%%%%%%%%%%%%%%%%%%%%%

	\begin{table}[t]
	\centering
	\caption{Assumed thruster operation condition \cite{Hofer:2007aa,ReidIEPC2015}.}
	\label{tab:condition}
	\begin{tabular}{p{45mm}p{15mm}}
		\Hline
		Parameter      & Value        \\
		\hline
		Channel centerline diameter  & 85 mm    \\
		Channel width  & 15 mm    \\
		Channel length  & 25 mm    \\
		Total mass flow rate  & 5.0 mg/s  \\
		Discharge voltage & 300 V   \\
		Anode temperature  & 850 K  \\
		\Hline
	\end{tabular}
	\end{table}
	
\section{1D Simulation in E-field direction}
	\label{sec:1d}
	\subsection{Calculation condition}
	
\begin{figure}[t]
		\begin{center}
			\includegraphics[width=65mm]{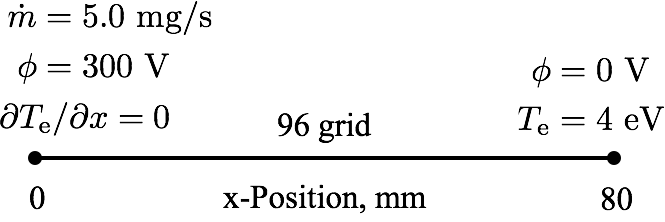}
		\end{center}
		\vspace{-4mm}
		\caption{Calculation domain and boundary conditions of the 1D simulation in the E-field (axial) direction.}
		\label{fig:condition1d}
	\end{figure}

	A 1D simulation in the E-field (axial) direction is conducted to validate the hybrid model. 
	The 1D simulation is performed by ignoring the distributions and gradients in the E$\times$B direction. 
	The calculation condition is shown in Fig. \ref{fig:condition1d}. 
	The calculation domain of 80 mm is taken in the axial direction of the Hall thruster. 
	All the xenon gas is introduced into the calculation domain from the anode side as neutral particles. 
	The velocity distribution of the neutral particles from the anode is determined by the Maxwellian distribution function of the anode temperature. 
	Discharge voltage is applied between the anode and cathode boundaries. 
	A Dirichlet condition of electron temperature is given at the cathode boundary whereas a Neumann condition is given at the anode side.
	
	\subsection{Results}
	
	A simulation is continued for several milli-seconds until a quasi-steady state is reached. 
	The simulated discharge current and thrust are 4.5 A and 108 mN, respectively. 
	The onboard operation of SPT-100 showed the discharge current and thrust of 4.6 A and 82 mN, respectively \cite{Manzella2001iepc}. 
	Therefore, the discharge current is well reproduced by the simulation whereas the thrust is overestimated by 32\%. 
	The overestimated thrust is attributed to the model assumption that the channel wall effects are ignored. 
	In the present simulation, all the ions are smoothly exhausted without colliding with the channel walls, yielding a large thrust.
	
\begin{figure}[t]
		\begin{center}
			\includegraphics[width=70mm]{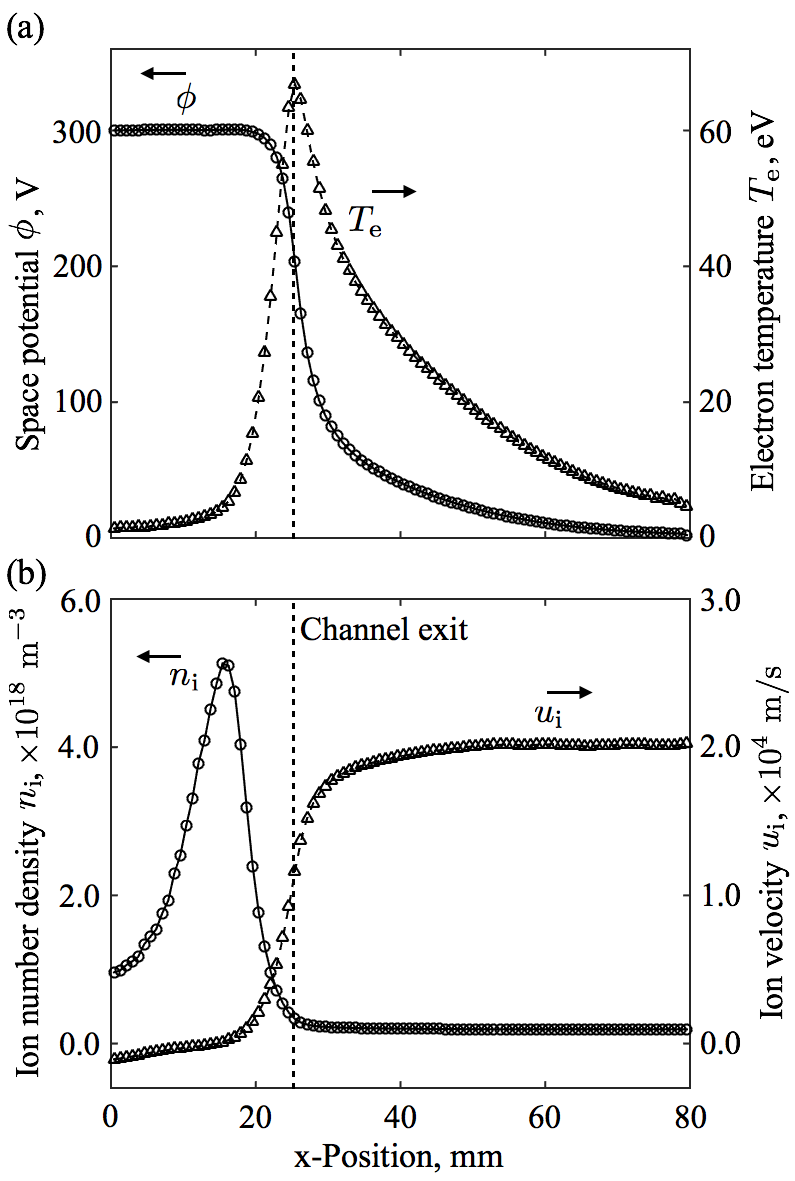}
		\end{center}
		\vspace{-4mm}
		\caption{Plasma property distributions simulated by the 1D model in the E-field direction. 
		(a) Space potential and electron temperature. (b) Ion number density and ion velocity.}
		\label{fig:dist1d}
	\end{figure}

	The plasma property distributions simulated by the 1D model are shown in Fig. \ref{fig:dist1d}. 
	The distribution tendencies of space potential $\phi$, electron temperature $T_{\rm e}$, and ion number density $n_{\rm i}$ are similar to the results of a previous numerical simulation \cite{Hofer:2007aa}, and the ion velocity distribution is consistent with the results of optical measurements \cite{Dorval:2002aa}. 
	A closer look at the results proves that the peak $n_{\rm i}$ and $T_{\rm e}$ are 5.1$\times$10$^{18}$ m$^{-3}$ and 67 eV, respectively. 
	These values are larger than the typical values observed in the previous simulation and experiment \cite{Hofer:2007aa,Dorval:2002aa}. 
	The reason of the overestimated $n_{\rm i}$ and $T_{\rm e}$ is again supposed to be the assumption of no channel walls. 
	By neglecting the channel walls, the ion and electron energy fluxes lost to the walls are assumed to be zero, resulting in the large $n_{\rm i}$ and $T_{\rm e}$. 
	Reflecting the channel wall effects to the models will be required to perform practical Hall thruster simulations.

	\begin{figure}[t]
		\begin{center}
			\includegraphics[width=65mm]{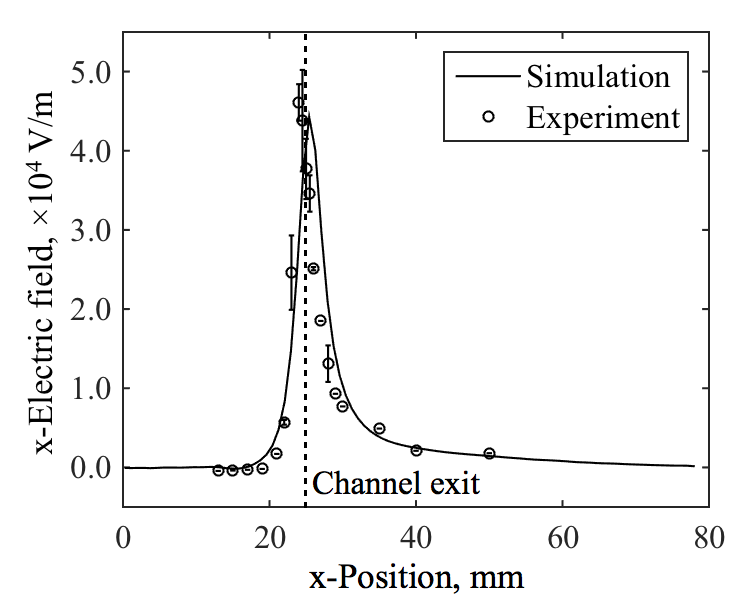}
		\end{center}
		\vspace{-4mm}
		\caption{Comparison of the electric field distributions obtained from the present simulation and experiment of the SPT-100 Hall thruster \cite{Dorval:2002aa}.}
		\label{fig:ex}
	\end{figure}
	
	The simulated electric field distribution is compared with experimental results \cite{Dorval:2002aa} in Fig. \ref{fig:ex}. 
	In both the numerical and experimental results, the main acceleration region exists at the vicinity of the channel exit. 
	The maximum electric fields obtained from the simulation and experiment are 4.4$\times$10$^{4}$ V/m and 4.6$\times$10$^{4}$ V/m, respectively, and the error is 4.1\%. 
	Hence the simulation result of electric field is considered to agree well with the experimental data. Since the potential is derived from the electron fluid model, this agreement indicates the validity of the present electron fluid model. 
	In summary, although the model assumption of the neglected channel walls remains an issue for realistic simulation of the thruster, the basic characteristics of the magnetized electron flow is well reflected in the simulation.

%\clearpage%%%%%%%%%%%%%%%%%%%%%%%%%%%%%%%%%%%%%%%%%%%%%%%%%%%%%%%%%%%%%%%%%%%%%%%%%%%%%%%

\section{2D Simulation in E-field and E$\times$B directions}
	\subsection{Calculation condition}
\begin{figure}[t]
		\begin{center}
			\includegraphics[width=60mm]{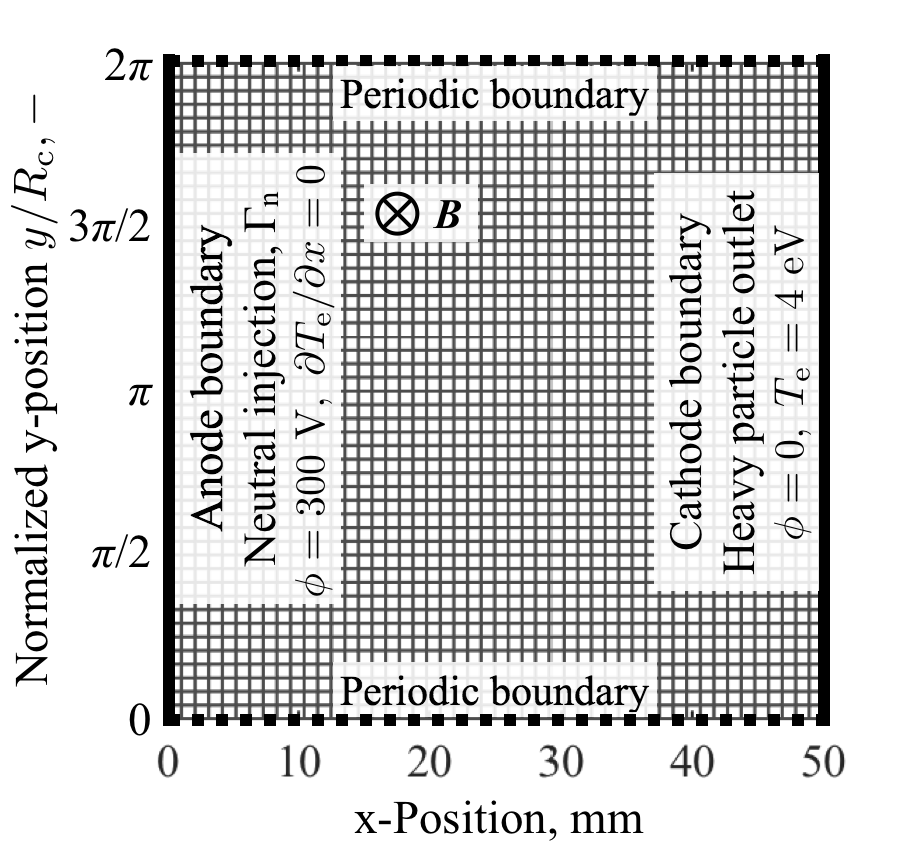}
		\end{center}
		\vspace{-4mm}
		\caption{Calculation region of the 2D simulation in the E-field and E$\times$B directions.}
		\label{fig:condition}
	\end{figure}

	The calculation domain and boundary conditions for the 2D simulation in the E-field and E$\times$B directions are shown in Fig. \ref{fig:condition}. 
	The full cylinder on the channel centerline is taken as the calculation domain. 
	The effects of finite channel curvature are neglected, and the Cartesian coordinate is used in this simulation. 
	The $y$-axis corresponds to the azimuthal direction and the $y$-position is normalized by the radius of the cylinder $R_{\rm c}$. 
	The boundary conditions for the anode and cathode sides are consistent with those used in the 1D simulation. 
	The periodic boundary condition is assumed on the top and bottom boundaries for both the particle and fluid calculations. 
	Referring the preceding axial-azimuthal simulation using a hybrid model \cite{Lam6922570}, a rectangular mesh with 48$\times$48 grid is used in the present simulation to obtain a good spatial resolution with a moderate computational cost. 
	The magnetic field is assumed to be directed to $-z$-direction and symmetric in $y$-direction. 
	The $x$-distribution of magnetic flux density is taken from the SPT-100 data \cite{Mitrofanova}.
	
\begin{figure}[t]
		\begin{center}
			\includegraphics[width=70mm]{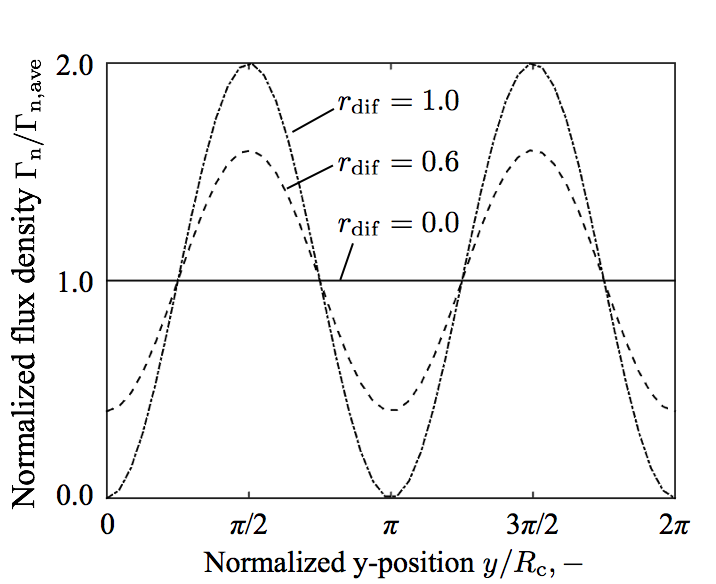}
		\end{center}
		\vspace{-4mm}
		\caption{Distribution of normalized neutral number flux density from the anode.}
		\label{fig:nflux}
	\end{figure}

	To add the nonuniformity in the mass flow density in the $y$-direction, the inflow neutral flux density from the anode $\Gamma_{\rm n}$ is controlled by cosine functions as shown in Fig. \ref{fig:nflux}. 
	Here the strength of azimuthal nonuniformity is represented by the differential mass flow ratio $r_{\rm dif}$ defined as follows: 
	\begin{equation}
		r_{\rm dif}=\frac{\Gamma_{\rm n,max}-\Gamma_{\rm n,min}}{2\Gamma_{\rm n,ave}},
	\end{equation}
	where $\Gamma_{\rm n,max}$, $\Gamma_{\rm n,min}$, and $\Gamma_{\rm n,ave}$ are the maximum, minimum, and average neutral flux density from the anode, respectively. 
	$\Gamma_{\rm n,max}$ is used at the $y/R_{\rm c}$ of $\pi/2$ and $3\pi/2$, and $\Gamma_{\rm n,min}$ is used at the $y/R_{\rm c}$ of 0, $\pi$, and $2\pi$. 
	In the case of $r_{\rm dif}$ = 0.0, the neutral particle flux density from the anode is uniform, whereas the case of $r_{\rm dif}$ = 1.0 gives the strongest azimuthal nonuniformity. 
	In the present study, $r_{\rm dif}$ is varied as 0.0, 0.6, and 1.0. 
	The $y$-velocity of inflow macroparticles is determined by the Maxwellian distribution function, whereas the $x$-velocity is given by the half-Maxwellian distribution with only positive values. 
	In other words, the velocity direction is random in the hemisphere in the $+x$-direction. 
	Owing to this assumption, the inflow neutral particles diffuse in the $y$-direction until they reach the main ionization region.
	
	
	\begin{figure*}[t]
		\begin{center}
			\includegraphics[width=120mm]{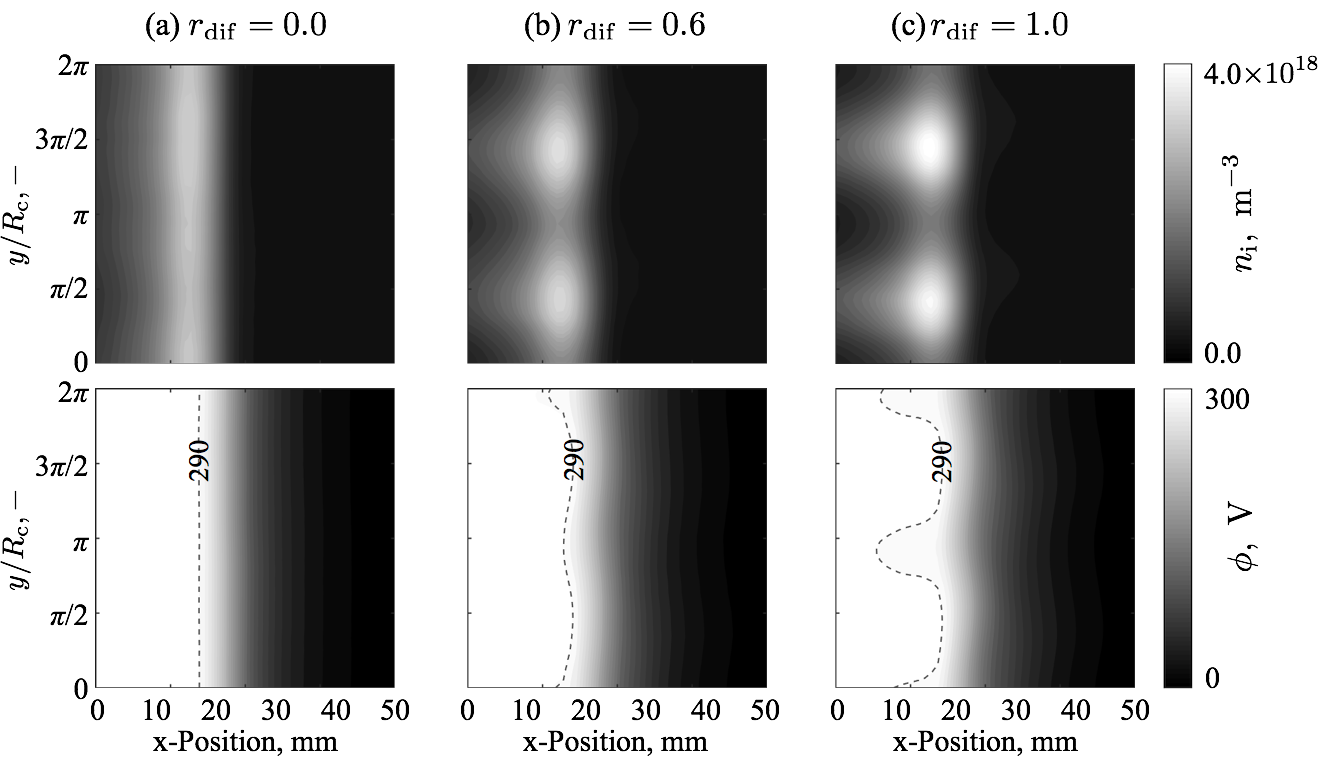}
		\end{center}
		\vspace{-4mm}
		\caption{Two-dimensional distributions of ion number density and space potential in the cases of (a) $r_{\rm dif}$ = 0.0, (b) $r_{\rm dif}$ = 0.6, and (c) $r_{\rm dif}$ = 1.0. 
		The equipotential lines of $\phi$ = 290 V are emphasized for better visualization.}
		\label{fig:dist2d}
	\end{figure*}

	\subsection{Two-dimensional plasma property distribution}
	
	The two-dimensional distributions of ion number density and space potential obtained by the present simulations are shown in Fig. \ref{fig:dist2d}. 
	In the case of uniform propellant supply ($r_{\rm dif}$ = 0.0), the distributions are uniform in $y$-direction. 
	The region of high ion number density is formed at around $x$ = 15 mm, as well as the result of the 1D simulation. On the other hand, in the cases of nonuniform propellant supply ($r_{\rm dif}$ = 0.6 and $r_{\rm dif}$ = 1.0), plasma property distributions become nonuniform in $y$-direction. 
	In the ion number density distributions, the nonuniformity becomes stronger as the differential mass flow ratio is increased. 
	In the results of nonuniform propellant supply cases, the peaks of ion number density exist at around the $y/R_{\rm c}$ of $\pi/2$ and $3\pi/2$, which correspond to the regions of high neutral particle flux densities from the anode. 
	High neutral particle flux causes large ionization rates, resulting in the high ion number density.
	
	The space potential distributions are also nonuniform in the $y$-direction in the cases of nonuniform propellant supply. 
	At around $x$ = 15 mm, high potential regions are observed at $y/R_{\rm c}$ of $\pi/2$ and $3\pi/2$, which correspond to the regions of high ion number density. 
	In the case of $r_{\rm dif}$ = 0.0, the electric field has only the $x$-component, and hence the direction of E$\times$B is essentially in the $y$-direction. 
	However, in the cases of nonuniform propellant supply, the azimuthal electric field $E_{y}$ exists and $E_{y}\times B_{z}$ effect in the $x$-direction is generated. 
	In the cylindrical coordinate of the annular Hall thruster, this effect corresponds to the $E_{\theta}\times B_{r}$ effect in the axial direction. 
	The electron motion is supposed to be affected by this $E_{\theta}\times B_{r}$ effect in the thruster operation of azimuthal nonuniform propellant supply.

	\subsection{Out-of-phase between $n_{\rm i}$ and $\phi$}

\begin{figure}[t]
		\begin{center}
			\includegraphics[width=63mm]{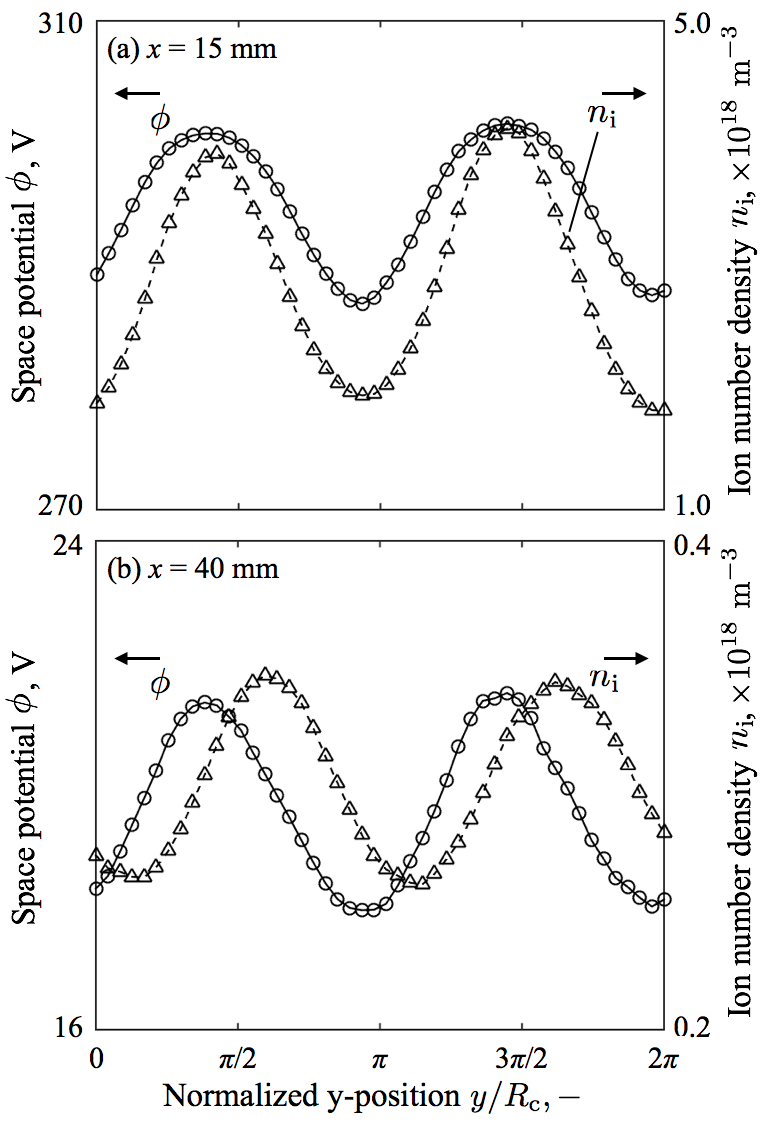}
		\end{center}
		\vspace{-4mm}
		\caption{Distributions of the ion number density and space potential in the $y$-direction at (a) $x$ = 15 mm and (b) $x$ = 40 mm.}
		\label{fig:dist_theta}
	\end{figure}

	To enhance the axial electron transport in Hall thrusters, the out-of-phase between the electron number density and space potential distributions is required, in addition to the presence of $E_{\theta}$. 
	Owing to the quasi-neutral assumption, the electron number density equates the ion number density $n_{\rm i}$. 
	The distributions in the $y$-direction of $n_{\rm i}$ and space potential $\phi$ of the $r_{\rm dif}$ = 1.0 case are shown in Fig. \ref{fig:dist_theta}. 
	At the upstream region ($x$ = 15 mm), the phases almost match between $n_{\rm i}$ and $\phi$. 
	In this case, the same amount of electron fluxes are induced to both $\pm x$ directions, and eventually they are cancelled out. 
	However, at the downstream region ($x$ = 40 mm), the phase shift is observed between $n_{\rm i}$ and $\phi$. 
	The high $n_{\rm i}$ regions exist in the azimuthal ranges of $\pi/2 < y/R_{\rm c} < 3\pi/4$ and $3\pi/2 < y/R_{\rm c} < 7\pi/4$, where $+E_y$ is generated. 
	On the other hand, at the low $n_{\rm i}$ regions, $-E_y$ is generated. 
	As shown in Fig. \ref{fig:condition}, the magnetic field is in the $-z$-direction. 
	Therefore, the $+E_y$ enhances the electron flux flowing to the upstream ($-x$) direction whereas the $-E_y$ causes an opposite effect. 
	Eventually, the net electron flux toward the anode is increased within the full azimuth. 
	One can consider that the azimuthally nonuniform propellant supply can enhance the axial cross-field electron transport.

	The mechanism for generating the out-of-phase has not been elucidated. 
	The key to understand this mechanism would be the electron transport throughout the discharge path. 
	The similar out-of-phase between the $n_{\rm i}$ and $\phi$ has also been reported in the rotating spokes, and its relationship with the axial electron transport has been concerned \cite{EllisonPoP2012}. 
	A detailed quantitative research on the effects of azimuthal nonuniformity on the cross-field electron transport is reserved for future study.

%\clearpage

\section{Conclusion}
	The effect of cross-field electron transport enhancement in the Hall thruster discharge with the azimuthally nonuniform propellant supply is qualitatively studied by a numerical simulation. 
	A two-dimensional hybrid model is used incorporated with an empirical anomalous electron mobility model. 
	The model is validated by the comparison of the axial electric fields obtained by the present simulation and the experiment. 
	Cosine functions of neutral flux density from the anode are assumed for the azimuthally differentiated propellant supply. 
	The main findings are summarized as follows:
	\begin{enumerate}
		\item Nonuniform plasma properties and the generation of electric field in the E$\times$B direction is confirmed.		
		\item Out-of-phase is observed in the E$\times$B direction between the distributions of electron number density and space potential.		
		\item Owing to the presence of the electric field and out-of-phase, the cross-field electron transport can be enhanced in the Hall thruster operation with azimuthally nonuniform propellant supply.
	\end{enumerate}

\section*{Acknowledgment}
   This work was supported by JSPS KAKENHI Grant Number JP17K14873.

%%\clearpage
\renewcommand{\refname}{Reference}
\bibliographystyle{elsarticle-num}
\bibliography{reference}

\end{document}